\documentclass[aps,prl,twocolumn,amsmath,amssymb,superscriptaddress,showpacs,floatfix]{revtex4-1}

\usepackage{graphicx}  	
\usepackage{float}  		
\usepackage{url}     		
\usepackage{longtable}    
\usepackage{threeparttable}
\usepackage{booktabs}
\usepackage{multirow}
\usepackage{dcolumn}
\usepackage{bm}
\usepackage{color}
\usepackage{amsmath}
\usepackage{xfrac}
\usepackage{verbatim}
\usepackage{array}
\usepackage{chngcntr}
\usepackage[colorlinks,hyperindex]{hyperref}
	\hypersetup
		{colorlinks,	
	 	citecolor=blue,	
	 	linkcolor=blue,	
	 	urlcolor=blue,	
		}
\graphicspath{{./}} 	
\begin{document}

\title{\textit{Ab-initio} Studies of (Li$_{0.8}$Fe$_{0.2}$)OHFeSe Superconductors: Revealing the Dual Roles of Fe$_{0.2}$ in Structural Stability and Charge Transfer}

\author{Wei Chen}
   \affiliation{International Center for Quantum Design of Functional Materials (ICQD), Hefei National Laboratory for Physical Sciences at Microscale, and Synergetic Innovation Center of Quantum Information and Quantum Physics, University of Science and Technology of China, Hefei, Anhui 230026, China}
   \affiliation{Department of Physics and School of Engineering and Applied Sciences, Harvard University, Cambridge, Massachusetts 02138, USA}
   
\author{Changgan Zeng}
   \affiliation{International Center for Quantum Design of Functional Materials (ICQD), Hefei National Laboratory for Physical Sciences at Microscale, and Synergetic Innovation Center of Quantum Information and Quantum Physics, University of Science and Technology of China, Hefei, Anhui 230026, China}

\author{Efthimios Kaxiras}
   \affiliation{Department of Physics and School of Engineering and Applied Sciences, Harvard University, Cambridge, Massachusetts 02138, USA}
     
\author{Zhenyu Zhang}
   \email{zhangzy@ustc.edu.cn}
   \affiliation{International Center for Quantum Design of Functional Materials (ICQD), Hefei National Laboratory for Physical Sciences at Microscale, and Synergetic Innovation Center of Quantum Information and Quantum Physics, University of Science and Technology of China, Hefei, Anhui 230026, China}

\date{\today}

\begin{abstract}
The recently discovered (Li$_{0.8}$Fe$_{0.2}$)OHFeSe superconductor provides a new platform for exploiting the microscopic mechanisms of high-$T_c$ superconductivity in FeSe-derived systems. Using density functional theory calculations, we first show that substitution of Li by Fe not only significantly strengthens the attraction between the (Li$_{0.8}$Fe$_{0.2}$)OH spacing layers and the FeSe superconducting layers along the \emph{c} axis, but also minimizes the lattice mismatch between the two in the \emph{ab} plane, both favorable for stabilizing the overall structure. Next we explore the electron injection into FeSe from the spacing layers, and unambiguously identify the Fe$_{0.2}$ components to be the dominant atomic origin of the dramatically enhanced interlayer charge transfer. We further reveal that the system strongly favors collinear antiferromagnetic ordering in the FeSe layers, but the spacing layers can be either antiferromagnetic or ferromagnetic depending on the Fe$_{0.2}$ spatial distribution. Based on these understandings, we also predict (Li$_{0.8}$Co$_{0.2}$)OHFeSe to be structurally stable with even larger electron injection and potentially higher $T_c$.
\end{abstract}


\maketitle

\newpage

Recently, two groups discovered (Li$_{1-x}$Fe$_{x}$)OHFeSe ($x\sim$0.2) as a new class of superconductors with high superconducting transition temperatures ($T_c$), and demonstrated coexistence of superconductivity and antiferromagnetism (AFM) or ferromagnetism (FM) in these systems~\cite{lu2014coexistence,pachmayr2015coexistence}. Such systems provide several appealing features for potentially revealing the likely superconducting mechanisms~\cite{lu2014coexistence,pachmayr2015coexistence,zhao2015common,niu2015surface}. However, despite quite a few subsequent experimental studies, limited knowledge has hitherto been obtained about the likely dominant roles played by the spacing layers~\cite{zhao2015common,niu2015surface,lei2015gate,yan2015surface,du2015two}, especially on how essential the dopant Fe$_{x}$ atoms are in establishing superconductivity. Intriguing questions include the magnetic ordering of the spacers, which appears to be inconsistent between the two pioneering experiments~\cite{lu2014coexistence,pachmayr2015coexistence}. More importantly, it is critically desirable to understand the structural stability of the systems formed by the weakly interacting spacers and superconducting FeSe monolayers~\cite{lu2014coexistence,pachmayr2015coexistence}, as well as the atomic origin of the electron doping from the (Li$_{1-x}$Fe$_{x}$)OH layers. Insights into those important issues, in turn, will be instrumental in searching for new FeSe-based superconductors with potentially higher $T_c$.

In this Letter, we use density functional theory (DFT) calculations to investigate the dominant roles of the spacers in establishing the high-$T_c$ superconductivity of (Li$_{0.8}$Fe$_{0.2}$)OHFeSe, with particular emphasis on the Fe$_{0.2}$ atoms. We find that substitution of Li by Fe strengthens the structural stability both in the \emph{ab} plane and along the \emph{c} axis. By further exploring the charge transfer, we identify the Fe$_{0.2}$ atoms to be the atomic origin of significant electron injection into FeSe. In addition, we obtain the ground-state magnetic order, and explain the seemingly controversial experimental observations on the magnetic ordering of the spacers. Based on these findings, we also predict a stable (Li$_{0.8}$Co$_{0.2}$)OHFeSe structure with larger electron doping into FeSe, potentially resulting in an even higher $T_c$.

\begin{figure}[ht]
	\begin{center}
		\includegraphics[width=3.1 in]{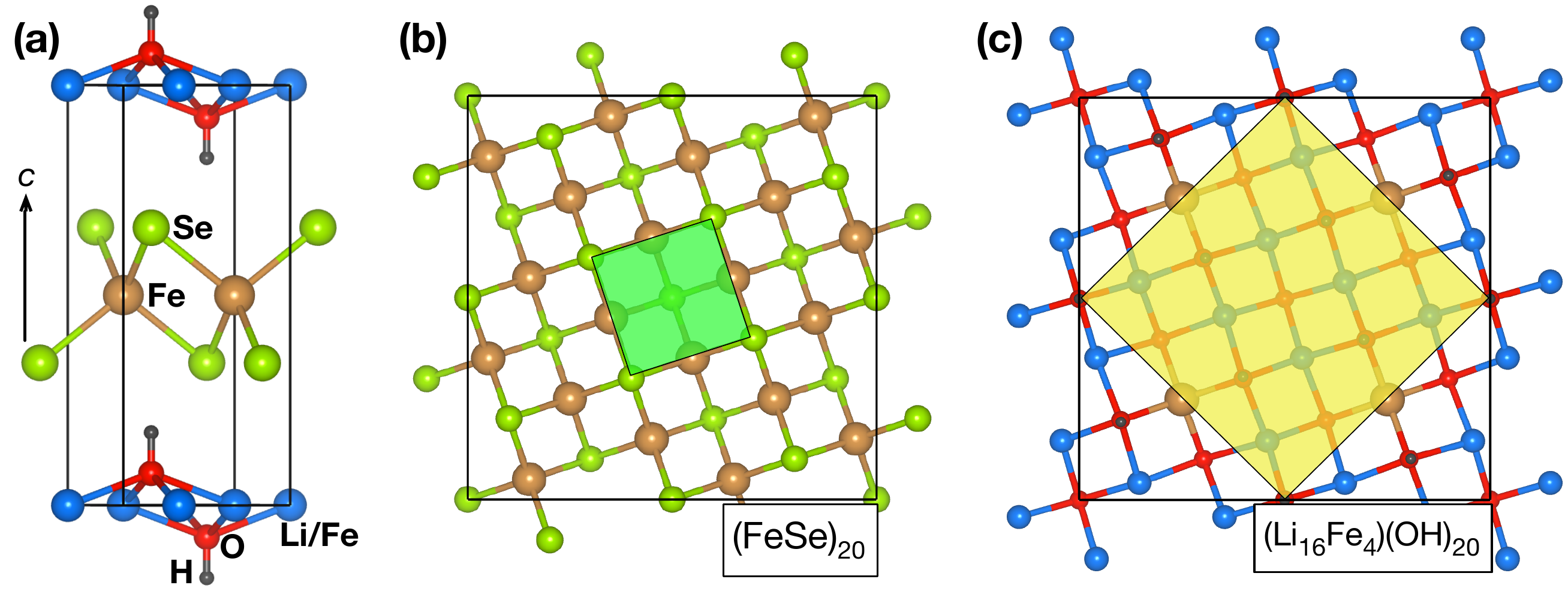}
	\end{center}
	\caption{(a) A side view of the $1\times1$ (Li/Fe)OHFeSe, and top views of (b) the FeSe and (c) (Li$_{0.8}$Fe$_{0.2}$)OH layer in a $\sqrt{10}\times\sqrt{10}$ cell. The green shaded area is the $1\times1$ cell, while the yellow is $\sqrt{5}\times\sqrt{5}$, which is compatible with states of nonmagnetic (NM), FM, or checkerboard AFM in FeSe. The collinear AFM in FeSe requires an expanded cell of $\sqrt{10}\times\sqrt{10}$.}
	\label{figure1}
\end{figure}

We first perform lattice-constant relaxation of (Li$_{0.8}$Fe$_{0.2}$)OHFeSe with different functionals and van der Waals (vdW) approaches, in order to choose the proper theoretical framework that can describe the current systems~\footnote{\label{note1}See Supplemental Materials for methods of DFT calculations, dependence of energy stability and magnetism on the Fe$_{0.2}$ distribution, calculations of magnetic coupling parameters, and several limitations of the present study}. In constructing the supercells, we have two considerations: (1) (Li$_{0.8}$Fe$_{0.2}$)OH layers have an occupation of Li/Fe with a ratio of $\sim$4~\cite{lu2014coexistence}, and thus the smallest cell is $\sqrt{5}\times\sqrt{5}$; (2) as discussed explicitly below, the ground-state magnetism is obtained to be collinear AFM for FeSe, consistent with several previous studies of FeSe/STO~\cite{zheng2013antiferromagnetic,PhysRevB.85.235123,PhysRevB.89.014501}. Therefore, a $\sqrt{10}\times\sqrt{10}$ cell is used to optimize the lattice parameters of ground-state (Li$_{0.8}$Fe$_{0.2}$)OHFeSe. Figs. \ref{figure1}b and \ref{figure1}c display the atomic structures of the FeSe and (Li$_{0.8}$Fe$_{0.2}$)OH layers; in a first study, the Fe$_{0.2}$ atoms in the spacer are orderly distributed and thus form a square lattice, where the nearest-neighboring Fe-Fe distance is a constant. Our detailed studies show that the clustering or disordering of the Fe$_{0.2}$ atoms is energetically less favorable~\cite{Note1}.

\begin{table}[ht]
\centering
\caption{Calculated lattice constants of (Li$_{0.8}$Fe$_{0.2}$)OHFeSe and LiOHFeSe. The vdW methods are based on the PBE functional. Numbers are in units of \AA. These results are obtained from optimization of the $\sqrt{10}\times\sqrt{10}$ cells, and the data is normalized to the $1\times1$ cells.}
\label{table1}
\scalebox{0.9}{
\begin{tabular}{|c|c|c|c|c|c|c|}
\hline
\multicolumn{6}{|c|}{(Li$_{0.8}$Fe$_{0.2}$)OHFeSe} & LiOHFeSe   \\ \hline
\multirow{2}{*}{EXP\footnote{EXP: experimental data at room temperature~\cite{lu2014coexistence}.}} & \multicolumn{2}{c|}{non-vdW} & \multicolumn{3}{c|}{vdW} & non-vdW \\ \cline{2-7} 
                     & LDA           & PBE          & DFT-D2      & DFT-TS      & vdW-DF2      & PBE       \\ \hline
\emph{a} = 3.786             & FTE\footnote{FTE: fix to the experimental parameter.\label{FTE}}             & 3.79            & 3.72      & 3.69      & FTE\textsuperscript{\ref{FTE}}      & 3.68       \\ \hline
\emph{c} = 9.288             & \textless\;8.20             & 9.38            & 8.52      & 8.30      & \textgreater\;9.78      &  11.16       \\ \hline
\end{tabular}
}
\end{table}

The optimized parameters are shown in Table~\ref{table1}, calculated from the energy dependence on the lattice constants. For LDA and vdW-DF2, we freeze \emph{a} to the experimental data~\cite{lu2014coexistence}, and find a large deviation in \emph{c}. We therefore relax only \emph{c} even without convergence for these two columns, while for the rest, both \emph{a} and \emph{c} are optimized until convergence is well reached. Between the non-vdW approaches, LDA significantly underestimates the lattice constant, while PBE slightly overestimates the axes, as commonly observed for the two functionals. The implemented vdW corrections as the empirical pairwise forms of ${C_{6}}/R^{6}_{0}$~\cite{JCC:JCC20495,PhysRevLett.102.073005} in DFT-D2 and DFT-TS are found to over-reduce \emph{c}; more surprisingly, the non-local vdW functional~\cite{PhysRevLett.92.246401,PhysRevB.82.081101,PhysRevB.83.195131} included in vdW-DF2 is found to severely overestimates \emph{c}. Here the failures of these vdW methods may originate from the excessively large $C_{6}$ coefficients due to the neglect of screening effect and many-body formalism for the former two, and from the imprecise exchange functional characterized in the latter. These problems call for in-depth investigations and possibly require major improvements on the methodology aspect. Presently, given the relatively small errors (\emph{a}: +0.16\%, \emph{c}: +0.96\%), we conclude that the PBE functional without including either of the popular vdW corrections yields a reasonable description, and thus the following results are all based within this framework.

To determine the roles of Fe$_{0.2}$ in the structural properties, we calculate the lattice constants of LiOHFeSe. The Fe atoms are removed from the spacer, and the collinear AFM order is found to remain for the FeSe layer. The obtained parameters (Table~\ref{table1}) are reduced by 3.01\% in \emph{a} and enlarged by 19.01\% in \emph{c} individually, from the lattice constants of (Li$_{0.8}$Fe$_{0.2}$)OHFeSe calculated by PBE. In the following, we analyze the effects on \emph{a} and \emph{c} separately in more details.

Experimentally, FeSe is inferred to be compressed in the \emph{ab} plane of (Li$_{0.8}$Fe$_{0.2}$)OHFeSe~\cite{lu2014coexistence}, where intuitively the strain should be caused by the relatively smaller lattice of (Li$_{0.8}$Fe$_{0.2}$)OH. Based on our results, the Fe$_{0.2}$ atoms indeed have expanded the \emph{ab} lattice; otherwise, the spacer could be even smaller in \emph{ab}, and the corresponding even larger mismatch with FeSe would make it unlikely to form a stable intercalated structure. To confirm this conjecture, we further calculate the lattice parameters of FeSe, (Li$_{0.8}$Fe$_{0.2}$)OH, and LiOH monolayers separately, whose \emph{a} values are found to be 3.74 \AA, 3.73 \AA, and 3.59 \AA, respectively. These values, together with that of bulk (Li$_{0.8}$Fe$_{0.2}$)OHFeSe, suggest that the in-plane compressed nature~\cite{lu2014coexistence} of FeSe may be inaccurate; however, our conjecture remains valid, given the small (\emph{or} large) lattice mismatch between FeSe and (Li$_{0.8}$Fe$_{0.2}$)OH (\emph{or} LiOH). Therefore, the Fe$_{0.2}$ atoms play a vital role in minimizing the lattice mismatch between the spacer and FeSe, which enables the formation of the \emph{commensurate stacking} as shown in Fig. \ref{figure1}a.

The \emph{FeSe and (Li$_{0.8}$Fe$_{0.2}$)OH layers are both in-plane stretched slightly} to 3.79 \AA\ when they are stacked alternatingly. This seems counter-intuitive, as the contacting interface generally adopts an intermediate lattice constant. To understand this observation, we examine the structures by calculating the thickness \emph{d} of each layer, defined as the distance from the upper Se \emph{or} H to the lower Se \emph{or} H position along the \emph{c} direction for the FeSe monolayer \emph{or} (Li$_{0.8}$Fe$_{0.2}$)OH spacing layer. We find \emph{d} is quenched from 3.48 to 3.33 \AA\ for (Li$_{0.8}$Fe$_{0.2}$)OH, while expanded from 2.89 to 2.91 \AA\ for FeSe, when the two layers are alternatively assembled into bulk. The volume expansion (from 3.74$^2$$\times$2.89 to 3.79$^2$$\times$2.91 \AA$^3$) of FeSe suggests an increased Coulomb repulsion internally, which \emph{strongly indicates electron injection}. The (Li$_{0.8}$Fe$_{0.2}$)OH layer is found to be rippled, with an amplitude of $\sim$0.12 \AA\ for H deviating from their average position along the \emph{c}-axis, making its in-plane lattice relatively easy to be stretched by interacting with the expanded FeSe, and the thickness tends to decrease as a compensation. Combining all the dimensions, the volume shrinks (from 3.73$^2$$\times$3.48 to 3.79$^2$$\times$3.33 \AA$^3$), supposed to be caused by electron depletion. To explain the counter-intuitive observation, here we emphasize the subtle correlation between lattice expansion and electron doping, which probably has been neglected in other FeSe-based systems (such as FeSe/STO and alkali-intercalated FeSe) that rely on charge transfer to enhance T$_c$.

We now examine specifically how Fe$_{0.2}$ influences the structures along the vertical directions. The \emph{c} axis is reduced from 11.16 \AA\ to 9.38 \AA\ when the Fe$_{0.2}$ atoms are incorporated into the spacer. Such a dramatic change suggests a greatly enhanced attraction between the spacer and FeSe. We thus calculate the cohesive energy: $E_C=E_{FeSe}+E_{Spacer}-E_{Bulk}$, where $E_{FeSe}$ is the total energy of the FeSe monolayer, $E_{Spacer}$ is that of (Li$_{0.8}$Fe$_{0.2}$)OH (\emph{or} LiOH), and $E_{Bulk}$ is that of the combined bulk. $E_{FeSe}$ and $E_{Spacer}$ are calculated using the in-plane lattice of the corresponding bulk system. Our results show that, $E_C$ per formula unit is 0.02 eV for LiOHFeSe and 0.35 eV for (Li$_{0.8}$Fe$_{0.2}$)OHFeSe. For the former, the value is even smaller than $E_C$ per carbon in graphite~\cite{PhysRevB.69.155406}, and the average bond distance $\overline{d}$ of the nearest Se-H is 3.50 \AA, clearly in the regime of weak vdW interaction. In contrast, for (Li$_{0.8}$Fe$_{0.2}$)OHFeSe, $E_C$ is stronger than that of vdW interaction, but still weaker than the typical strength of chemical bonding. This moderate interaction originated from the incorporated Fe$_{0.2}$ substantially stabilizes the structure vertically, while still allows mechanical cleavage of the crystal~\cite{niu2015surface,yan2015surface,du2015two}. More importantly, such an increased $E_C$ and the closer contact ($\overline{d}$ = 3.11 \AA) between FeSe and (Li$_{0.8}$Fe$_{0.2}$)OH collectively enable enhanced charge transfer between the layers. 

\begin{figure}[ht]
	\begin{center}
		\includegraphics[width=3.3 in]{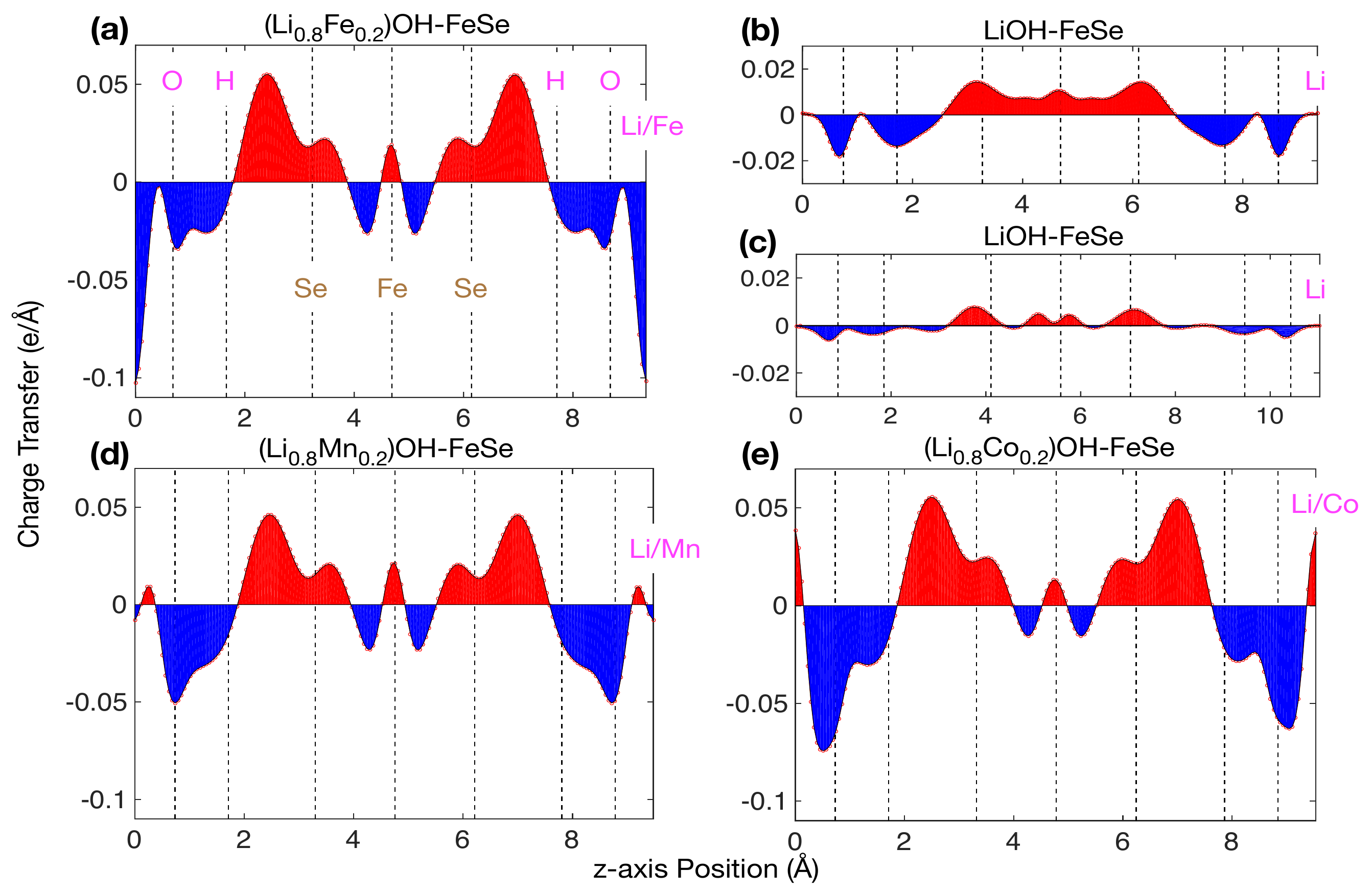}
	\end{center}
	\caption{$\Delta\rho_{z}$ of (a) (Li$_{0.8}$Fe$_{0.2}$)OHFeSe and (b,c) LiOHFeSe at the lattice constant of (Li$_{0.8}$Fe$_{0.2}$)OHFeSe and of LiOHFeSe, (d) (Li$_{0.8}$Mn$_{0.2}$)OHFeSe, and (e) (Li$_{0.8}$Co$_{0.2}$)OHFeSe at their own relaxed lattices. These results are calculated from the ground states of $\sqrt{10}\times\sqrt{10}$ cells, and have been normalized to $1\times1$ cells. The red and blue shaded areas indicate electron accumulation and depletion, respectively.}
	\label{figure2}
\end{figure}

Following the above structural indications, we now investigate the detailed nature of charge transfer. We calculate the charge density difference $\Delta\rho$ between the combined (Li$_{0.8}$Fe$_{0.2}$)OHFeSe bulk system and the sum of the isolated FeSe and (Li$_{0.8}$Fe$_{0.2}$)OH layers. To have a quantitative picture, we plot the plane-averaged $\Delta\rho$ along the \emph{c} axis ($\Delta\rho_{z}$) in Fig.~\ref{figure2}a. To understand the roles of Fe$_{0.2}$, we also calculate $\Delta\rho_{z}$ for LiOHFeSe at the fixed lattice constant of (Li$_{0.8}$Fe$_{0.2}$)OHFeSe and at the relaxed lattice, as shown in Figs.~\ref{figure2}b and~\ref{figure2}c respectively. 

We find that, when the (Li$_{0.8}$Fe$_{0.2}$)OH and FeSe layers merge in bulk, the charge density between Fe and Se in FeSe has an electron-depletion region, and this amount of charge is mostly transferred to the Se planes to form interactions with the spacers. Such features are not observed in LiOHFeSe, where the interlayer $E_C$ is much weaker. In each plot, we shift the average \emph{z} position of the Fe atoms in FeSe to be located exactly at the center of the supercell. The charge redistribution is mirror symmetric with respect to the middle line in Figs.~\ref{figure2}a and~\ref{figure2}b. However, in Fig.~\ref{figure2}c, the layers are loosely stacked with the interlayer distance found to be slightly alternating (the distance between the Se and H planes differs by $\sim$0.15 \AA\ at the two opposite sides of FeSe), making the curve not to be precisely symmetric. To see the doping levels of FeSe, we integrate $\Delta\rho_{z}$ over \emph{z} within the Se boundary on each side. We define a boundary between H and Se, so that the distance from Se (H) to the boundary is proportional to the atomic radius of Se (H)~\cite{jcp/41/10/10.1063/1.1725697}. The electron injection $\rho_{I}$ into FeSe is calculated to be 0.051 electrons per FeSe in (Li$_{0.8}$Fe$_{0.2}$)OHFeSe. For LiOHFeSe, $\rho_{I}$ = 0.015$e/Fe$ and 0.005$e/Fe$ for structures with the (Li$_{0.8}$Fe$_{0.2}$)OHFeSe and LiOHFeSe lattice constant, respectively. The estimated value of 0.051 is close to the experimentally measured value of $\sim$0.08 $e/Fe$~\cite{zhao2015common}, and the difference might be due to the absence of structural defects or disordering of Fe$_{0.2}$ in the supercell systems.

From a broader perspective, a sufficiently large electron doping is essential to realize high $T_c$ in FeSe-based superconductors. In the FeSe/STO systems, O vacancies in STO are believed to be the atomic origin of large interlayer charge transfer~\cite{PhysRevB.87.220503}. Here in (Li$_{0.8}$Fe$_{0.2}$)OHFeSe, one may naturally think that the atomic origin of electron injection is the Fe$_{0.2}$ atoms in the spacer, because of their generally higher oxidation state than Li. Indeed, our results agree with this conjecture, and demonstrate that the Fe$_{0.2}$ atoms \emph{enhance the electron injection into FeSe in two ways}. First, by comparing Figs.~\ref{figure2}a and~\ref{figure2}b where the lattice constants are identical, we clearly see the large contribution of the Fe$_{0.2}$ \emph{d} orbitals to the charge transfer when Li is substituted. Second, the increased doping in Fig.~\ref{figure2}b compared to Fig.~\ref{figure2}c indicates that the closer interlayer coupling caused by Fe$_{0.2}$ also boosts the amount of charge transfer. These two effects are both crucial, and should be instrumental in search for other novel spacer-intercalated superconductors.

The interlayer charge transfer results in the higher $T_c$ in bulk (Li$_{0.8}$Fe$_{0.2}$)OHFeSe than in bulk FeSe; however, the doping level of FeSe in (Li$_{0.8}$Fe$_{0.2}$)OHFeSe is still lower than that in FeSe/STO~\cite{zhao2015common}, suggesting the feasibility of other structural design to further enhance $T_c$. Based on the above studies of the roles of Fe$_{0.2}$, we can attempt to substitute Li by other elements X instead of Fe in the spacers, to see if such elements can also stabilize the structure, and more importantly, induce larger charge transfer. Presently, we have examined X = Mn or Co, which potentially has higher oxidation state than that of Fe. Our calculations show that both (Li$_{0.8}$Mn$_{0.2}$)OHFeSe and (Li$_{0.8}$Co$_{0.2}$)OHFeSe are indeed structurally stable, with their lattice constants similar to that of (Li$_{0.8}$Fe$_{0.2}$)OHFeSe. Furthermore, based on the charge transfer shown in Figs.~\ref{figure2}d and~\ref{figure2}e, $\rho_{I}$ is calculated to be 0.044$e/Fe$ for (Li$_{0.8}$Mn$_{0.2}$)OHFeSe and 0.060$e/Fe$ for (Li$_{0.8}$Co$_{0.2}$)OHFeSe. The Mn and Co atoms do not appear to contribute to the charge injection of FeSe as much as Fe, but significantly facilitate more electron transfer from the O atoms in the spacer. The depleted area in FeSe of (Li$_{0.8}$Co$_{0.2}$)OHFeSe is also found to be smaller than that in (Li$_{0.8}$Fe$_{0.2}$)OHFeSe. The larger $\rho_{I}$ of (Li$_{0.8}$Co$_{0.2}$)OHFeSe suggests a higher $T_c$ than that of (Li$_{0.8}$Fe$_{0.2}$)OHFeSe. More studies should be made along this line of structural design, involving other possible elements and substitution concentrations $>$ 0.2, to further identify the best candidates in this family of materials to realize much higher $T_c$. 

\begin{figure}[ht]
	\begin{center}
		\includegraphics[width=3.1 in]{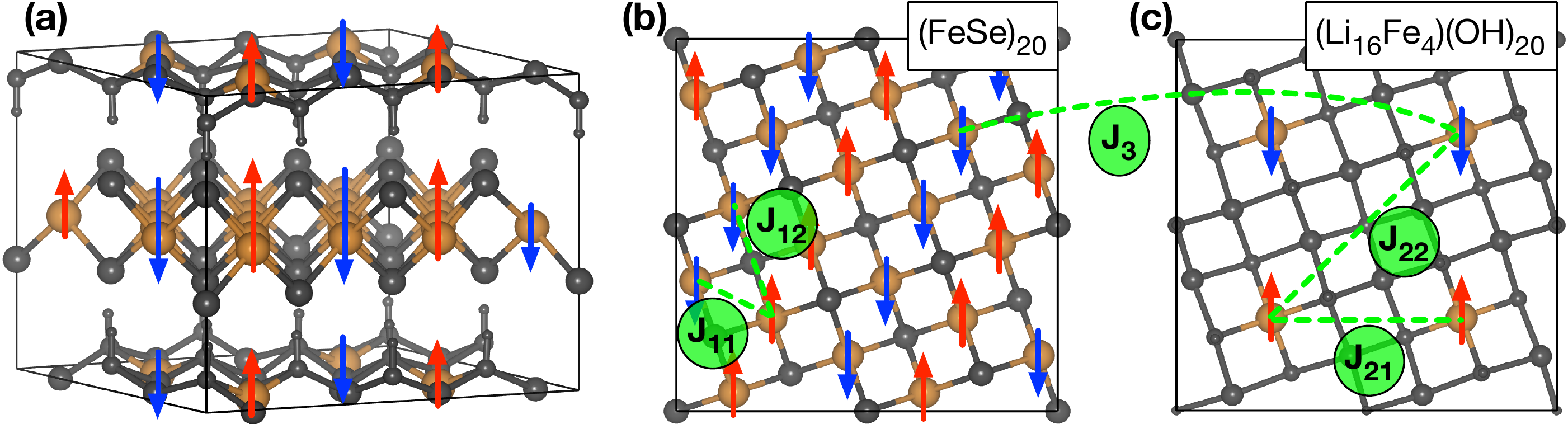}
	\end{center}
	\caption{(a) Ground-state magnetism of (Li$_{0.8}$Fe$_{0.2}$)OHFeSe in the (b) FeSe and (c) (Li$_{0.8}$Fe$_{0.2}$)OH layers. Elements other than Fe have moments of $\sim$0, and are colored in gray. In (a), the arrows spanning over the Fe rows indicate the same spin orientation in the row. The green lines in (b) and (c) connect the nearest or next-nearest Fe neighbors, whose magnetic couplings are considered in the Heisenberg model.}
	\label{figure3}
\end{figure}

We next focus on the magnetic properties, as magnetism is generally related to superconductivity in the FeSe-derived systems. We examine four different magnetic orders in each of the FeSe and (Li$_{0.8}$Fe$_{0.2}$)OH layers, including NM, FM, checkerboard AFM, and collinear AFM, and investigate their possible combinations for interlayer coupling. Our calculations show that FM is unable to be established in FeSe, as well as NM in (Li$_{0.8}$Fe$_{0.2}$)OH. The computed magnetic moment of Fe$_{0.2}$ is about 3.53$\mu_{B}$, and for that of Fe in FeSe, it is $\sim$ 2.25$\mu_{B}$ in the collinear AFM and $\sim$ 1.86$\mu_{B}$ in the checkerboard AFM state. The larger moment of Fe$_{0.2}$ suggests that the spin magnitude is reduced in a closer packing of the Fe atoms in FeSe. By comparing the total energies of the different orders, we obtain the \emph{magnetic ground state of (Li$_{0.8}$Fe$_{0.2}$)OHFeSe} (Fig.~\ref{figure3}). Both layers exhibit a collinear AFM order in the Fe square lattices, and the interlayer spins are aligned parallel to each other. AFM is calculated to be slightly more stable than FM by only 2 meV per Fe in (Li$_{0.8}$Fe$_{0.2}$)OH. Such results of the fragile AFM ground state in the spacer also help to explain the recent experimental observations~\cite{wu2015nmr}. In addition, for the newly designed systems, (Li$_{0.8}$Co$_{0.2}$)OHFeSe has the same spin configuration as (Li$_{0.8}$Fe$_{0.2}$)OHFeSe, while (Li$_{0.8}$Mn$_{0.2}$)OHFeSe exhibits anti-parallel alignment between the collinear AFM FeSe and (Li$_{0.8}$Mn$_{0.2}$)OH layers.

Based on these DFT inputs, we quantitatively analyze the coupling strengths in (Li$_{0.8}$Fe$_{0.2}$)OHFeSe using the Heisenberg model on square lattices~\cite{PhysRevLett.102.177003,PhysRevB.89.014501,PhysRevB.91.020504}. We use an approximate Hamiltonian:
\begin{equation} 
	\label{equation1}
	\footnotesize{H \hspace{-0.09cm}=\hspace{-0.08cm} [\hspace{-0.02cm}(J_{\text{\tiny{11}}}\hspace{-0.22cm}\sum_{\textless ij\textgreater}\hspace{-0.15cm}+J_{\text{\tiny{12}}}\hspace{-0.24cm}\sum_{\ll ij\gg}\hspace{-0.14cm})\vec{S_i}\hspace{-0.06cm}\cdot\hspace{-0.06cm}\vec{S_j}\hspace{-0.02cm}]_{\text{\tiny{1}}}
	\hspace{-0.10cm}+\hspace{-0.08cm}[\hspace{-0.02cm}(J_{\text{\tiny{21}}}\hspace{-0.20cm}\sum_{\textless ij\textgreater}\hspace{-0.15cm}+J_{\text{\tiny{22}}}\hspace{-0.24cm}\sum_{\ll ij\gg}\hspace{-0.14cm})\vec{S_i}\hspace{-0.06cm}\cdot\hspace{-0.06cm}\vec{S_j}\hspace{-0.02cm}]_{\text{\tiny{2}}}
	\hspace{-0.08cm}+\hspace{-0.08cm}[\hspace{-0.02cm}J_{\text{\tiny{3}}}\hspace{-0.20cm}\sum_{\textless ij\textgreater}\hspace{-0.14cm}\vec{S_i}\hspace{-0.06cm}\cdot\hspace{-0.06cm}\vec{S_j}\hspace{-0.02cm}]_{\text{\tiny{3}}}}.
\end{equation}
The three terms respectively account for couplings in FeSe, in (Li$_{0.8}$Fe$_{0.2}$)OH, and between the two layers. $\textless ij\textgreater$ and $\ll\hspace{-0.1cm} ij\hspace{-0.1cm}\gg$ represent summations over the nearest and next-nearest neighbors. Based on the energies of different magnetic states, we estimate the coupling strengths (Table~\ref{table2})~\cite{Note1}. We note that the computed $J_{11}$ and $J_{12}$ in FeSe are quite close to previous results~\cite{PhysRevLett.102.177003,PhysRevB.89.014501}.

\begin{table}[ht]
\centering
\caption{Calculated $J$ values of (Li$_{0.8}$Fe$_{0.2}$)OHFeSe, in which the Fe$_{0.2}$ atoms are uniformly distributed in spacer.}
\small
\label{table2}
\scalebox{1.0}{
\begin{tabular}{|c|c|c|c|c|c|}
\hline
 & $J_{11}$ & $J_{12}$ & $J_{21}$ & $J_{22}$ & $J_{3}$ \\ \hline
Unit: meV & 83.18 & 47.82 & 0.05 & $-0.02$ & $-1.94$ \\ \hline
\end{tabular}
}
\end{table}

Generally speaking, the order of the coupling strength is: $J_{11,12}\gg J_{3} \gg J_{21,22}$, just the reverse of the Fe-Fe pairing distance ($d_{11,12}< d_{3}< d_{21,22}$). The couplings in (Li$_{0.8}$Fe$_{0.2}$)OH are quite weak, making its magnetism more dominated by the relatively larger coupling between the layers. The magnetism of FeSe and the negative $J_3$ thus result in a collinear AFM ground state in (Li$_{0.8}$Fe$_{0.2}$)OH (Fig.~\ref{figure3}c), with the spin direction of each site parallel to that of the corresponding Fe in FeSe. Furthermore, the much weaker $J_{3}$ compared to $J_{11}$ and $J_{12}$ suggests that, Fe$_{0.2}$ should play a minimal role in directly influencing the magnetism of FeSe by magnetic interplay between the layers. This fact also indicates that the magnetic coupling between FeSe and the spacer is \emph{unlikely to play an important role, or probably even undesirable} in establishing high $T_c$.

It is notable that, the above ground-state magnetism is computed using an ordered distribution of Fe$_{0.2}$ in the spacer. In the actual systems, certain degrees of disordering is unavoidable. This structural fact could possibly change the coupling strengths in the spacer and the Fe sites in FeSe that are coupled closely to the Fe$_{0.2}$ atoms. By calculating the ground-state magnetism, here we demonstrate that disordering of Fe$_{0.2}$ can possibly result in a FM order in the spacer~\cite{Note1}. This finding may help to clarify the controversial experimental observations of different magnetic orders in (Li$_{0.8}$Fe$_{0.2}$)OH~\cite{lu2014coexistence,pachmayr2015coexistence}.

Overall, despite several limitations~\cite{Note1}, this study has revealed the dual roles of Fe$_{0.2}$ in the structural stability and electronic charge injection into FeSe. Such roles are critically important in the fabrication of FeSe-based high-$T_c$ superconductors, and may provide new insights into exploration of the likely pairing mechanisms. Our predicted (Li$_{0.8}$Co$_{0.2}$)OH superconductors with larger charge transfer and potentially higher $T_c$ also calls for experimental fabrication and validation.

\bibliography{ref} 	

\begin{thebibliography}{22}%
\makeatletter
\providecommand \@ifxundefined [1]{%
 \@ifx{#1\undefined}
}%
\providecommand \@ifnum [1]{%
 \ifnum #1\expandafter \@firstoftwo
 \else \expandafter \@secondoftwo
 \fi
}%
\providecommand \@ifx [1]{%
 \ifx #1\expandafter \@firstoftwo
 \else \expandafter \@secondoftwo
 \fi
}%
\providecommand \natexlab [1]{#1}%
\providecommand \enquote  [1]{``#1''}%
\providecommand \bibnamefont  [1]{#1}%
\providecommand \bibfnamefont [1]{#1}%
\providecommand \citenamefont [1]{#1}%
\providecommand \href@noop [0]{\@secondoftwo}%
\providecommand \href [0]{\begingroup \@sanitize@url \@href}%
\providecommand \@href[1]{\@@startlink{#1}\@@href}%
\providecommand \@@href[1]{\endgroup#1\@@endlink}%
\providecommand \@sanitize@url [0]{\catcode `\\12\catcode `\$12\catcode
  `\&12\catcode `\#12\catcode `\^12\catcode `\_12\catcode `\%12\relax}%
\providecommand \@@startlink[1]{}%
\providecommand \@@endlink[0]{}%
\providecommand \url  [0]{\begingroup\@sanitize@url \@url }%
\providecommand \@url [1]{\endgroup\@href {#1}{\urlprefix }}%
\providecommand \urlprefix  [0]{URL }%
\providecommand \Eprint [0]{\href }%
\providecommand \doibase [0]{http://dx.doi.org/}%
\providecommand \selectlanguage [0]{\@gobble}%
\providecommand \bibinfo  [0]{\@secondoftwo}%
\providecommand \bibfield  [0]{\@secondoftwo}%
\providecommand \translation [1]{[#1]}%
\providecommand \BibitemOpen [0]{}%
\providecommand \bibitemStop [0]{}%
\providecommand \bibitemNoStop [0]{.\EOS\space}%
\providecommand \EOS [0]{\spacefactor3000\relax}%
\providecommand \BibitemShut  [1]{\csname bibitem#1\endcsname}%
\let\auto@bib@innerbib\@empty
\bibitem [{\citenamefont {Lu}\ \emph {et~al.}(2015)\citenamefont {Lu} \emph
  {et~al.}}]{lu2014coexistence}%
  \BibitemOpen
  \bibfield  {author} {\bibinfo {author} {\bibfnamefont {X.~F.}\ \bibnamefont
  {Lu}} \emph {et~al.},\ }\href
  {http://www.nature.com/nmat/journal/v14/n3/full/nmat4155.html} {\bibfield
  {journal} {\bibinfo  {journal} {Nat. Mater.}\ }\textbf {\bibinfo {volume}
  {14}},\ \bibinfo {pages} {325} (\bibinfo {year} {2015})}\BibitemShut
  {NoStop}%
\bibitem [{\citenamefont {Pachmayr}\ \emph {et~al.}(2015)\citenamefont
  {Pachmayr}, \citenamefont {Nitsche}, \citenamefont {Luetkens}, \citenamefont
  {Kamusella}, \citenamefont {Br{\"u}ckner}, \citenamefont {Sarkar},
  \citenamefont {Klauss},\ and\ \citenamefont
  {Johrendt}}]{pachmayr2015coexistence}%
  \BibitemOpen
  \bibfield  {author} {\bibinfo {author} {\bibfnamefont {U.}~\bibnamefont
  {Pachmayr}}, \bibinfo {author} {\bibfnamefont {F.}~\bibnamefont {Nitsche}},
  \bibinfo {author} {\bibfnamefont {H.}~\bibnamefont {Luetkens}}, \bibinfo
  {author} {\bibfnamefont {S.}~\bibnamefont {Kamusella}}, \bibinfo {author}
  {\bibfnamefont {F.}~\bibnamefont {Br{\"u}ckner}}, \bibinfo {author}
  {\bibfnamefont {R.}~\bibnamefont {Sarkar}}, \bibinfo {author} {\bibfnamefont
  {H.-H.}\ \bibnamefont {Klauss}}, \ and\ \bibinfo {author} {\bibfnamefont
  {D.}~\bibnamefont {Johrendt}},\ }\href
  {http://onlinelibrary.wiley.com/doi/10.1002/anie.201407756/full} {\bibfield
  {journal} {\bibinfo  {journal} {Angew. Chem. Int. Ed.}\ }\textbf {\bibinfo
  {volume} {54}},\ \bibinfo {pages} {293} (\bibinfo {year} {2015})}\BibitemShut
  {NoStop}%
\bibitem [{\citenamefont {Zhao}\ \emph {et~al.}(2015)\citenamefont {Zhao} \emph
  {et~al.}}]{zhao2015common}%
  \BibitemOpen
  \bibfield  {author} {\bibinfo {author} {\bibfnamefont {L.}~\bibnamefont
  {Zhao}} \emph {et~al.},\ }\href {http://arxiv.org/abs/1505.06361} {\bibfield
  {journal} {\bibinfo  {journal} {arXiv:1505.06361}\ } (\bibinfo {year}
  {2015})}\BibitemShut {NoStop}%
\bibitem [{\citenamefont {Niu}\ \emph {et~al.}(2015)\citenamefont {Niu} \emph
  {et~al.}}]{niu2015surface}%
  \BibitemOpen
  \bibfield  {author} {\bibinfo {author} {\bibfnamefont {X.~H.}\ \bibnamefont
  {Niu}} \emph {et~al.},\ }\href {\doibase 10.1103/PhysRevB.92.060504}
  {\bibfield  {journal} {\bibinfo  {journal} {Phys. Rev. B}\ }\textbf {\bibinfo
  {volume} {92}},\ \bibinfo {pages} {060504} (\bibinfo {year}
  {2015})}\BibitemShut {NoStop}%
\bibitem [{\citenamefont {Lei}\ \emph {et~al.}(2015)\citenamefont {Lei},
  \citenamefont {Xiang}, \citenamefont {Lu}, \citenamefont {Wang},
  \citenamefont {Chang}, \citenamefont {Shang}, \citenamefont {Luo},
  \citenamefont {Wu}, \citenamefont {Sun},\ and\ \citenamefont
  {Chen}}]{lei2015gate}%
  \BibitemOpen
  \bibfield  {author} {\bibinfo {author} {\bibfnamefont {B.}~\bibnamefont
  {Lei}}, \bibinfo {author} {\bibfnamefont {Z.~J.}\ \bibnamefont {Xiang}},
  \bibinfo {author} {\bibfnamefont {X.~F.}\ \bibnamefont {Lu}}, \bibinfo
  {author} {\bibfnamefont {N.~Z.}\ \bibnamefont {Wang}}, \bibinfo {author}
  {\bibfnamefont {J.~R.}\ \bibnamefont {Chang}}, \bibinfo {author}
  {\bibfnamefont {C.}~\bibnamefont {Shang}}, \bibinfo {author} {\bibfnamefont
  {X.~G.}\ \bibnamefont {Luo}}, \bibinfo {author} {\bibfnamefont
  {T.}~\bibnamefont {Wu}}, \bibinfo {author} {\bibfnamefont {Z.}~\bibnamefont
  {Sun}}, \ and\ \bibinfo {author} {\bibfnamefont {X.~H.}\ \bibnamefont
  {Chen}},\ }\href {http://arxiv.org/abs/1503.02457} {\bibfield  {journal}
  {\bibinfo  {journal} {arXiv:1503.02457}\ } (\bibinfo {year}
  {2015})}\BibitemShut {NoStop}%
\bibitem [{\citenamefont {Yan}\ \emph {et~al.}(2015)\citenamefont {Yan} \emph
  {et~al.}}]{yan2015surface}%
  \BibitemOpen
  \bibfield  {author} {\bibinfo {author} {\bibfnamefont {Y.~J.}\ \bibnamefont
  {Yan}} \emph {et~al.},\ }\href {http://arxiv.org/abs/1507.02577} {\bibfield
  {journal} {\bibinfo  {journal} {arXiv:1507.02577}\ } (\bibinfo {year}
  {2015})}\BibitemShut {NoStop}%
\bibitem [{\citenamefont {Du}\ \emph {et~al.}(2015)\citenamefont {Du},
  \citenamefont {Yang}, \citenamefont {Lin}, \citenamefont {Fang},
  \citenamefont {Du}, \citenamefont {Xing}, \citenamefont {Yang}, \citenamefont
  {Zhu},\ and\ \citenamefont {Wen}}]{du2015two}%
  \BibitemOpen
  \bibfield  {author} {\bibinfo {author} {\bibfnamefont {Z.}~\bibnamefont
  {Du}}, \bibinfo {author} {\bibfnamefont {X.}~\bibnamefont {Yang}}, \bibinfo
  {author} {\bibfnamefont {H.}~\bibnamefont {Lin}}, \bibinfo {author}
  {\bibfnamefont {D.}~\bibnamefont {Fang}}, \bibinfo {author} {\bibfnamefont
  {G.}~\bibnamefont {Du}}, \bibinfo {author} {\bibfnamefont {J.}~\bibnamefont
  {Xing}}, \bibinfo {author} {\bibfnamefont {H.}~\bibnamefont {Yang}}, \bibinfo
  {author} {\bibfnamefont {X.}~\bibnamefont {Zhu}}, \ and\ \bibinfo {author}
  {\bibfnamefont {H.-H.}\ \bibnamefont {Wen}},\ }\href
  {http://arxiv.org/abs/1506.04645} {\bibfield  {journal} {\bibinfo  {journal}
  {arXiv:1506.04645}\ } (\bibinfo {year} {2015})}\BibitemShut {NoStop}%
\bibitem [{Note1()}]{Note1}%
  \BibitemOpen
  \bibinfo {note} {\label {note1}See Supplemental Materials for methods of DFT
  calculations, dependence of energy stability and magnetism on the Fe$_{0.2}$
  distribution, calculations of magnetic coupling parameters, and several
  limitations of the present study}\BibitemShut {NoStop}%
\bibitem [{\citenamefont {Zheng}\ \emph {et~al.}(2013)\citenamefont {Zheng},
  \citenamefont {Wang}, \citenamefont {Kang},\ and\ \citenamefont
  {Zhang}}]{zheng2013antiferromagnetic}%
  \BibitemOpen
  \bibfield  {author} {\bibinfo {author} {\bibfnamefont {F.}~\bibnamefont
  {Zheng}}, \bibinfo {author} {\bibfnamefont {Z.}~\bibnamefont {Wang}},
  \bibinfo {author} {\bibfnamefont {W.}~\bibnamefont {Kang}}, \ and\ \bibinfo
  {author} {\bibfnamefont {P.}~\bibnamefont {Zhang}},\ }\href
  {http://www.nature.com/srep/2013/130718/srep02213/full/srep02213.html}
  {\bibfield  {journal} {\bibinfo  {journal} {Sci. Rep.}\ }\textbf {\bibinfo
  {volume} {3}},\ \bibinfo {pages} {2213} (\bibinfo {year} {2013})}\BibitemShut
  {NoStop}%
\bibitem [{\citenamefont {Liu}\ \emph {et~al.}(2012)\citenamefont {Liu},
  \citenamefont {Lu},\ and\ \citenamefont {Xiang}}]{PhysRevB.85.235123}%
  \BibitemOpen
  \bibfield  {author} {\bibinfo {author} {\bibfnamefont {K.}~\bibnamefont
  {Liu}}, \bibinfo {author} {\bibfnamefont {Z.-Y.}\ \bibnamefont {Lu}}, \ and\
  \bibinfo {author} {\bibfnamefont {T.}~\bibnamefont {Xiang}},\ }\href
  {\doibase 10.1103/PhysRevB.85.235123} {\bibfield  {journal} {\bibinfo
  {journal} {Phys. Rev. B}\ }\textbf {\bibinfo {volume} {85}},\ \bibinfo
  {pages} {235123} (\bibinfo {year} {2012})}\BibitemShut {NoStop}%
\bibitem [{\citenamefont {Cao}\ \emph {et~al.}(2014)\citenamefont {Cao},
  \citenamefont {Tan}, \citenamefont {Xiang}, \citenamefont {Feng},\ and\
  \citenamefont {Gong}}]{PhysRevB.89.014501}%
  \BibitemOpen
  \bibfield  {author} {\bibinfo {author} {\bibfnamefont {H.-Y.}\ \bibnamefont
  {Cao}}, \bibinfo {author} {\bibfnamefont {S.}~\bibnamefont {Tan}}, \bibinfo
  {author} {\bibfnamefont {H.}~\bibnamefont {Xiang}}, \bibinfo {author}
  {\bibfnamefont {D.~L.}\ \bibnamefont {Feng}}, \ and\ \bibinfo {author}
  {\bibfnamefont {X.-G.}\ \bibnamefont {Gong}},\ }\href {\doibase
  10.1103/PhysRevB.89.014501} {\bibfield  {journal} {\bibinfo  {journal} {Phys.
  Rev. B}\ }\textbf {\bibinfo {volume} {89}},\ \bibinfo {pages} {014501}
  (\bibinfo {year} {2014})}\BibitemShut {NoStop}%
\bibitem [{\citenamefont {Grimme}(2006)}]{JCC:JCC20495}%
  \BibitemOpen
  \bibfield  {author} {\bibinfo {author} {\bibfnamefont {S.}~\bibnamefont
  {Grimme}},\ }\href {\doibase 10.1002/jcc.20495} {\bibfield  {journal}
  {\bibinfo  {journal} {J. Comput. Chem.}\ }\textbf {\bibinfo {volume} {27}},\
  \bibinfo {pages} {1787} (\bibinfo {year} {2006})}\BibitemShut {NoStop}%
\bibitem [{\citenamefont {Tkatchenko}\ and\ \citenamefont
  {Scheffler}(2009)}]{PhysRevLett.102.073005}%
  \BibitemOpen
  \bibfield  {author} {\bibinfo {author} {\bibfnamefont {A.}~\bibnamefont
  {Tkatchenko}}\ and\ \bibinfo {author} {\bibfnamefont {M.}~\bibnamefont
  {Scheffler}},\ }\href {\doibase 10.1103/PhysRevLett.102.073005} {\bibfield
  {journal} {\bibinfo  {journal} {Phys. Rev. Lett.}\ }\textbf {\bibinfo
  {volume} {102}},\ \bibinfo {pages} {073005} (\bibinfo {year}
  {2009})}\BibitemShut {NoStop}%
\bibitem [{\citenamefont {Dion}\ \emph {et~al.}(2004)\citenamefont {Dion},
  \citenamefont {Rydberg}, \citenamefont {Schr\"oder}, \citenamefont
  {Langreth},\ and\ \citenamefont {Lundqvist}}]{PhysRevLett.92.246401}%
  \BibitemOpen
  \bibfield  {author} {\bibinfo {author} {\bibfnamefont {M.}~\bibnamefont
  {Dion}}, \bibinfo {author} {\bibfnamefont {H.}~\bibnamefont {Rydberg}},
  \bibinfo {author} {\bibfnamefont {E.}~\bibnamefont {Schr\"oder}}, \bibinfo
  {author} {\bibfnamefont {D.~C.}\ \bibnamefont {Langreth}}, \ and\ \bibinfo
  {author} {\bibfnamefont {B.~I.}\ \bibnamefont {Lundqvist}},\ }\href {\doibase
  10.1103/PhysRevLett.92.246401} {\bibfield  {journal} {\bibinfo  {journal}
  {Phys. Rev. Lett.}\ }\textbf {\bibinfo {volume} {92}},\ \bibinfo {pages}
  {246401} (\bibinfo {year} {2004})}\BibitemShut {NoStop}%
\bibitem [{\citenamefont {Lee}\ \emph {et~al.}(2010)\citenamefont {Lee},
  \citenamefont {Murray}, \citenamefont {Kong}, \citenamefont {Lundqvist},\
  and\ \citenamefont {Langreth}}]{PhysRevB.82.081101}%
  \BibitemOpen
  \bibfield  {author} {\bibinfo {author} {\bibfnamefont {K.}~\bibnamefont
  {Lee}}, \bibinfo {author} {\bibfnamefont {E.~D.}\ \bibnamefont {Murray}},
  \bibinfo {author} {\bibfnamefont {L.}~\bibnamefont {Kong}}, \bibinfo {author}
  {\bibfnamefont {B.~I.}\ \bibnamefont {Lundqvist}}, \ and\ \bibinfo {author}
  {\bibfnamefont {D.~C.}\ \bibnamefont {Langreth}},\ }\href {\doibase
  10.1103/PhysRevB.82.081101} {\bibfield  {journal} {\bibinfo  {journal} {Phys.
  Rev. B}\ }\textbf {\bibinfo {volume} {82}},\ \bibinfo {pages} {081101}
  (\bibinfo {year} {2010})}\BibitemShut {NoStop}%
\bibitem [{\citenamefont {Klime\ifmmode~\check{s}\else \v{s}\fi{}}\ \emph
  {et~al.}(2011)\citenamefont {Klime\ifmmode~\check{s}\else \v{s}\fi{}},
  \citenamefont {Bowler},\ and\ \citenamefont
  {Michaelides}}]{PhysRevB.83.195131}%
  \BibitemOpen
  \bibfield  {author} {\bibinfo {author} {\bibfnamefont {J.}~\bibnamefont
  {Klime\ifmmode~\check{s}\else \v{s}\fi{}}}, \bibinfo {author} {\bibfnamefont
  {D.~R.}\ \bibnamefont {Bowler}}, \ and\ \bibinfo {author} {\bibfnamefont
  {A.}~\bibnamefont {Michaelides}},\ }\href {\doibase
  10.1103/PhysRevB.83.195131} {\bibfield  {journal} {\bibinfo  {journal} {Phys.
  Rev. B}\ }\textbf {\bibinfo {volume} {83}},\ \bibinfo {pages} {195131}
  (\bibinfo {year} {2011})}\BibitemShut {NoStop}%
\bibitem [{\citenamefont {Zacharia}\ \emph {et~al.}(2004)\citenamefont
  {Zacharia}, \citenamefont {Ulbricht},\ and\ \citenamefont
  {Hertel}}]{PhysRevB.69.155406}%
  \BibitemOpen
  \bibfield  {author} {\bibinfo {author} {\bibfnamefont {R.}~\bibnamefont
  {Zacharia}}, \bibinfo {author} {\bibfnamefont {H.}~\bibnamefont {Ulbricht}},
  \ and\ \bibinfo {author} {\bibfnamefont {T.}~\bibnamefont {Hertel}},\ }\href
  {\doibase 10.1103/PhysRevB.69.155406} {\bibfield  {journal} {\bibinfo
  {journal} {Phys. Rev. B}\ }\textbf {\bibinfo {volume} {69}},\ \bibinfo
  {pages} {155406} (\bibinfo {year} {2004})}\BibitemShut {NoStop}%
\bibitem [{\citenamefont {Slater}(1964)}]{jcp/41/10/10.1063/1.1725697}%
  \BibitemOpen
  \bibfield  {author} {\bibinfo {author} {\bibfnamefont {J.~C.}\ \bibnamefont
  {Slater}},\ }\href {\doibase http://dx.doi.org/10.1063/1.1725697} {\bibfield
  {journal} {\bibinfo  {journal} {J. Chem. Phys.}\ }\textbf {\bibinfo {volume}
  {41}},\ \bibinfo {pages} {3199} (\bibinfo {year} {1964})}\BibitemShut
  {NoStop}%
\bibitem [{\citenamefont {Bang}\ \emph {et~al.}(2013)\citenamefont {Bang} \emph
  {et~al.}}]{PhysRevB.87.220503}%
  \BibitemOpen
  \bibfield  {author} {\bibinfo {author} {\bibfnamefont {J.}~\bibnamefont
  {Bang}} \emph {et~al.},\ }\href {\doibase 10.1103/PhysRevB.87.220503}
  {\bibfield  {journal} {\bibinfo  {journal} {Phys. Rev. B}\ }\textbf {\bibinfo
  {volume} {87}},\ \bibinfo {pages} {220503} (\bibinfo {year}
  {2013})}\BibitemShut {NoStop}%
\bibitem [{\citenamefont {Wu}\ \emph {et~al.}(2015)\citenamefont {Wu},
  \citenamefont {Zhao}, \citenamefont {Lian}, \citenamefont {Lu}, \citenamefont
  {Wang}, \citenamefont {Luo}, \citenamefont {Chen},\ and\ \citenamefont
  {Wu}}]{wu2015nmr}%
  \BibitemOpen
  \bibfield  {author} {\bibinfo {author} {\bibfnamefont {Y.~P.}\ \bibnamefont
  {Wu}}, \bibinfo {author} {\bibfnamefont {D.}~\bibnamefont {Zhao}}, \bibinfo
  {author} {\bibfnamefont {X.~R.}\ \bibnamefont {Lian}}, \bibinfo {author}
  {\bibfnamefont {X.~F.}\ \bibnamefont {Lu}}, \bibinfo {author} {\bibfnamefont
  {N.~Z.}\ \bibnamefont {Wang}}, \bibinfo {author} {\bibfnamefont {X.~G.}\
  \bibnamefont {Luo}}, \bibinfo {author} {\bibfnamefont {X.~H.}\ \bibnamefont
  {Chen}}, \ and\ \bibinfo {author} {\bibfnamefont {T.}~\bibnamefont {Wu}},\
  }\href {http://journals.aps.org/prb/abstract/10.1103/PhysRevB.91.125107}
  {\bibfield  {journal} {\bibinfo  {journal} {Phys. Rev. B}\ }\textbf {\bibinfo
  {volume} {91}},\ \bibinfo {pages} {125107} (\bibinfo {year}
  {2015})}\BibitemShut {NoStop}%
\bibitem [{\citenamefont {Ma}\ \emph {et~al.}(2009)\citenamefont {Ma},
  \citenamefont {Ji}, \citenamefont {Hu}, \citenamefont {Lu},\ and\
  \citenamefont {Xiang}}]{PhysRevLett.102.177003}%
  \BibitemOpen
  \bibfield  {author} {\bibinfo {author} {\bibfnamefont {F.}~\bibnamefont
  {Ma}}, \bibinfo {author} {\bibfnamefont {W.}~\bibnamefont {Ji}}, \bibinfo
  {author} {\bibfnamefont {J.}~\bibnamefont {Hu}}, \bibinfo {author}
  {\bibfnamefont {Z.-Y.}\ \bibnamefont {Lu}}, \ and\ \bibinfo {author}
  {\bibfnamefont {T.}~\bibnamefont {Xiang}},\ }\href {\doibase
  10.1103/PhysRevLett.102.177003} {\bibfield  {journal} {\bibinfo  {journal}
  {Phys. Rev. Lett.}\ }\textbf {\bibinfo {volume} {102}},\ \bibinfo {pages}
  {177003} (\bibinfo {year} {2009})}\BibitemShut {NoStop}%
\bibitem [{\citenamefont {Cao}\ \emph {et~al.}(2015)\citenamefont {Cao},
  \citenamefont {Chen}, \citenamefont {Xiang},\ and\ \citenamefont
  {Gong}}]{PhysRevB.91.020504}%
  \BibitemOpen
  \bibfield  {author} {\bibinfo {author} {\bibfnamefont {H.-Y.}\ \bibnamefont
  {Cao}}, \bibinfo {author} {\bibfnamefont {S.}~\bibnamefont {Chen}}, \bibinfo
  {author} {\bibfnamefont {H.}~\bibnamefont {Xiang}}, \ and\ \bibinfo {author}
  {\bibfnamefont {X.-G.}\ \bibnamefont {Gong}},\ }\href {\doibase
  10.1103/PhysRevB.91.020504} {\bibfield  {journal} {\bibinfo  {journal} {Phys.
  Rev. B}\ }\textbf {\bibinfo {volume} {91}},\ \bibinfo {pages} {020504}
  (\bibinfo {year} {2015})}\BibitemShut {NoStop}%
\end{thebibliography}%

\end{document}